\documentclass[twocolumn]{aastex631}

\newcommand{\kms}{km~s$^{-1}$}

\newcommand{\msun}{$M_{\odot}$}

\shorttitle{Anisotropy of Halo Main Sequence Turnoff Stars}
\shortauthors{King et al.}

\begin{document}

\title{Anisotropy of Halo Main Sequence Turnoff Stars Measured with New MMT Radial Velocities and Gaia Proper Motions}


\author[0000-0001-8665-3376]{Charles King III}
\affil{Smithsonian Astrophysical Observatory, 60 Garden St, Cambridge, MA 02138, USA}

\author[0000-0002-4462-2341]{Warren R. Brown}
\affil{Smithsonian Astrophysical Observatory, 60 Garden St, Cambridge, MA 02138, USA}

\author[0000-0002-9146-4876]{Margaret J. Geller}
\affil{Smithsonian Astrophysical Observatory, 60 Garden St, Cambridge, MA 02138, USA}

\author[0000-0003-0214-609X]{Scott J. Kenyon}
\affil{Smithsonian Astrophysical Observatory, 60 Garden St, Cambridge, MA 02138, USA}

\begin{abstract}

We measure the anisotropy of the Milky Way stellar halo traced by a dense sample of $18<r<21$ mag F-type main sequence turnoff stars using Gaia eDR3 proper motions and new radial velocity measurements published here.

\end{abstract}


\section{Introduction}

The motions of stars are determined by the Milky Way's gravitational potential and provide a measure of its mass distribution.  Many tools exist to analyze stellar velocity distributions.  The anisotropy parameter, $\beta$, is a summary statistic of the ratio of tangential to radial random motion \citep{binney80}.  A system of stars with purely radial orbits has $\beta=1$; a system with purely circular orbits has $\beta=-\infty$.  Theoretical simulations demonstrate that changes in $\beta$ in the stellar halo can be a sensitive probe of both recent interactions with satellites and ancient major mergers \citep{loebman18}.  

Our previous work, based on radial velocities of F-type main sequence turnoff stars in the halo, found evidence for a surprising dip in anisotropy around $R\sim20$ kpc \citep{king15} similar to that found by \citet{kafle12} for blue horizontal branch stars.  Gaia now provides precise proper motions for these stars.  Gaia-based studies of nearby solar neighborhood stars \citep{helmi18} and distant blue horizontal branch (BHB) stars \citep{belokurov18} conclude that the Milky Way has a two-component inner halo, and that radially anisotropic stars arrived from a major $10^{11}$ \msun\ merger event around 10 Gyr ago. Main sequence turnoff stars are less luminous than evolved or giant stars, but provide a much denser tracer at $20$ kpc distances.

We revisit our original result using Gaia proper motions and new radial velocity measurements presented here.  

\section{Measurements}

As described in \citet{king15}, we select $18<r<21$ mag F-type stars using extinction-corrected colors from the Sloan Digital Sky Survey.  We obtain spectra for 4,022 new stars at the 6.5m MMT telescope with the Hectospec spectrograph between 2015 September and 2017 April, of which 2,851 satisfy our S/N$>$4 per pixel quality threshold.  We measure radial velocities using the RVSAO cross-correlation package.  The median uncertainty is 18 \kms\ at the median $r=20$ mag.

We also make use of F-type stars published in the SDSS Stellar Parameter Pipeline catalog \citep{allende08}.  This catalog is relatively shallow; the entire catalog is brighter than the median $r=20$ mag of the Hectospec sample.  

We obtain proper motions and their covariances from Gaia eDR3 \citep{lindegren21a}.  We impose quality thresholds recommended by the Gaia Collaboration, \texttt{ruwe} $<$ 1.4 being the most important.  The mean proper motion error at $r=20$ mag corresponds to a tangential velocity error $\pm30$ \kms\ if distance error is zero.  Parallax provides no meaningful constraint for our stars, so we estimate distances using the photometric parallax relation of \citet{ivezic08} and \citet{bond10}.  Statistical distance errors are 15\%.  

We transform measurements to the Galactic coordinate frame assuming the Sun is located at $(X,Y,Z)=(-8,0,0.02)$ kpc, the local circular velocity is 235 \kms, and the local standard of rest motion of \citet{schonrich10}.  We then impose $|Z|>$ 5 kpc to remove all significant disk contamination, leaving us with a sample of 17,221 halo F-type stars.  We note that the Sgr stream lies beyond our sample depth, and does not enter into our sample.

\section{Results}

\begin{figure*}
 \centerline{\includegraphics[width=5.5in]{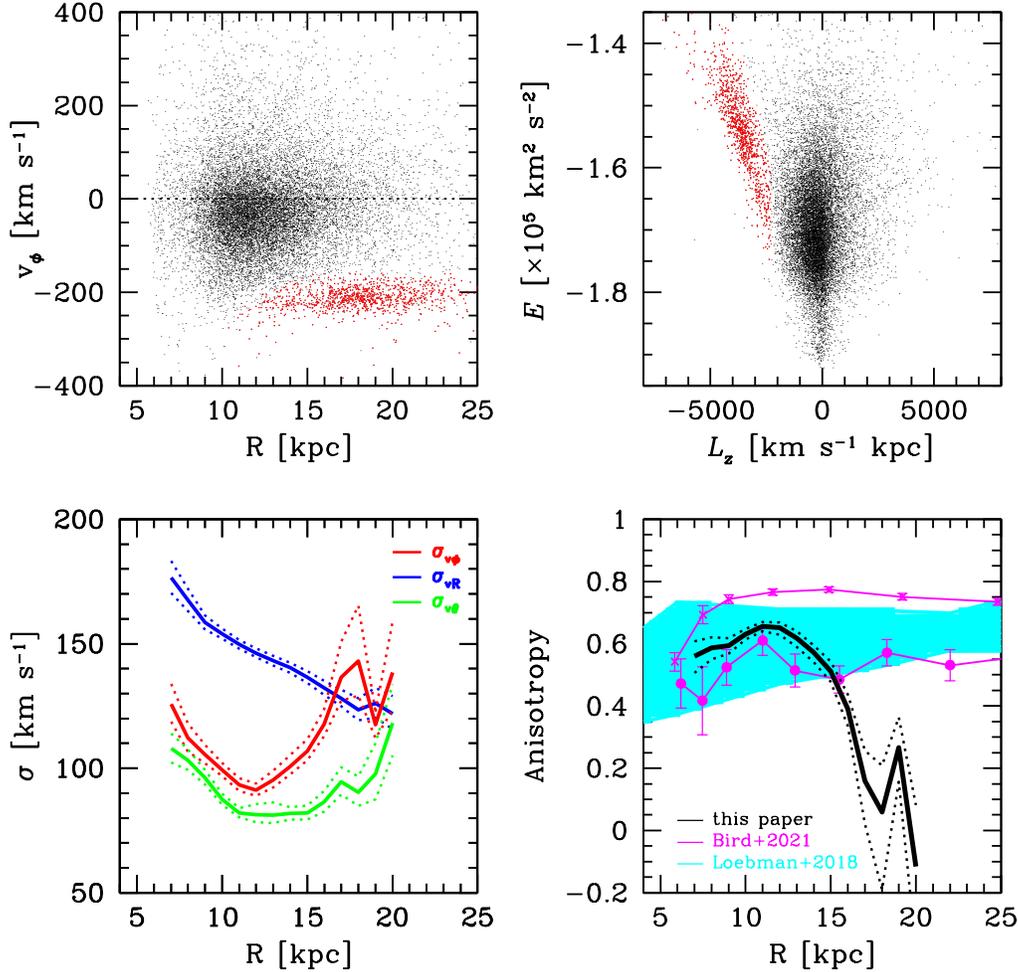}}
 \caption{  \label{fig:results}
	Azimuthal velocities of our halo F-type stars (upper left panel) reveal a thin disk-like population contaminating our sample at $15\lesssim R \lesssim 20$ kpc that we associate with ACS/Mon (red dots).  The population is easily seen in energy $E$ angular momentum $L_Z$ space (upper right panel) and affects the computed velocity dispersions $\sigma$ (lower left panel) and anisotropy (lower right panel) of our full sample.  We compare our anisotropy (black line) with theoretical predictions \citep[cyan band,][]{loebman18} and observed K giants (magenta crosses) and BHB stars (magenta circles) \citep{bird21}.	Dotted lines plot 1$\sigma$ uncertainties. New radial velocities are available as Data Behind the Figure.
	}
\end{figure*}

The measured distributions of velocity components are similar to our previous work with one exception:  we find an unexpected group of stars with thin disk azimuthal velocity $v_{\phi}\simeq-230$ \kms\ that dominates our sample by number around $R\sim18$ kpc (Figure 1).  These stars are located in the Galactic anti-center $100\arcdeg<l<208\arcdeg$ at $b<55\arcdeg$ latitudes.  This is the ACS/Mon halo structure \citep{xu15, morganson16}.  We consider orbital energy $E$ and angular momentum $L_Z$ per unit mass using the gravitational potential of \citet{kenyon14}, and find that the ACS/Mon stars have $L_Z<-2300$ \kms~kpc$^{-1}$ within $8L_Z -1.95 \times 10^5 < E < 15L_Z -1.95 \times 10^5$ km$^2$~s$^{-2}$ (red dots in Figure 1).

We compute mean velocities and velocity dispersions by partitioning the full sample into 2 kpc wide bins in Galactocentric radius $R$.  We estimate errors from bootstrap re-sampling $10^4$ times using the full covariance matrix of observational errors.  The mean radial and longitudinal velocities are consistent with zero, as expected, although the mean azimuthal velocity, $<v_{\phi}>=-25$ \kms, is slightly prograde.  The velocity covariances are consistent with zero.  The tilt angles of the velocity ellipsoid are closely aligned with the spherical coordinate system.  

The lower panels of Figure 1 plot the velocity dispersions and the anisotropy of our full sample.  In the range $15\lesssim R \lesssim 20$ kpc, the ACS/Mon stars artificially increase the computed azimuthal velocity dispersion $\sigma_{\phi}$, and decrease the anisotropy (Figure 1).  Other samples that include stars in the Galactic anti-center region with $b<55\arcdeg$ may be similarly affected, depending on the relative number of stars from that region of sky.

In the range $7<R<15$ kpc, the mean anisotropy, $\beta\simeq0.6$, of our halo F-type stars is consistent with BHB stars (lower magenta line) but is more tangential than the $\beta$ measured for K giants (upper magenta line) \citep{bird21}.  For comparison, \citet{loebman18} compute theoretical $\beta$ profiles for various accretion-only and cosmological hydrodynamic simulations. All of the simulations predict a radial anisotropy $<$$\beta$$>$ $\sim 0.6$ (cyan band) consistent with our result for $7<R<15$ kpc.

We provide the 2,851 new radial velocities as Data Behind the Figure.


\begin{acknowledgments}
We thank Perry Berlind and Michael Calkins for their assistance with observations obtained at the MMT Observatory, a joint facility of the Smithsonian Institution and the University of Arizona.  We thank Sarah Bird and Alison Deason for their correspondence. This work makes use of data from the European Space Agency Gaia mission, data from the Sloan Digital Sky Survey, and data products created by the Smithsonian Astrophysical Observatory OIR Telescope Data Center.  This research also makes use of NASA's Astrophysics Data System Bibliographic Services.  This work was supported by the Smithsonian Institution. 
\end{acknowledgments}

\facilities{MMT (Hectospec Spectrograph), Gaia}

\end{document}